\documentstyle[nato,graphicx,amsmath]{crckapb}

\newcommand{\romb}{{\operatorname{b}}}
\newcommand{\romB}{{\operatorname{B}}}
\newcommand{\romc}{{\operatorname{c}}}
\newcommand{\romd}{{\operatorname{d}}}
\newcommand{\romD}{{\operatorname{D}}}
\newcommand{\rome}{{\operatorname{e}}}
\newcommand{\romf}{{\operatorname{f}}}
\newcommand{\romGC}{{\operatorname{GC}}}
\newcommand{\romi}{{\operatorname{i}}}
\newcommand{\romid}{{\operatorname{id}}}
\newcommand{\romM}{{\operatorname{M}}}
\newcommand{\romP}{{\operatorname{P}}}
\newcommand{\romPB}{{\operatorname{PB}}}
\newcommand{\romr}{{\operatorname{r}}}
\newcommand{\roms}{{\operatorname{s}}}
\newcommand{\romT}{{\operatorname{T}}}
\newcommand{\romtot}{{\operatorname{tot}}}
\newcommand{\romTr}{{\operatorname{Tr}}}

\newcommand{\VECp}{{\boldsymbol{p}}}
\newcommand{\VECr}{{\boldsymbol{r}}}

\newcommand{\ie}{$i$.$\,e$.}
\newcommand{\eg}{$e$.$\,g$.}
\newcommand{\etal}{{$et.\,al$}}

\begin{opening}

\title{Cell model and Poisson-Boltzmann theory:\protect\\ A brief introduction}
\author{Markus Deserno$^1$,
        Christian Holm$^2$}
\institute{$^1$ Department of Chemistry and Biochemistry, UCLA, USA\\
           $^2$ Max-Planck-Institut f\"ur Polymerforschung, Mainz, Germany\\
Email: $^1${\tt markus@chem.ucla.edu}, $^2${\tt holm@mpip-mainz.mpg.de}}
\end{opening}

\begin{document}




The cell-model\index{Cell-model} and its treatment on the Poisson-Boltzmann\index{Poisson-Boltzmann} level are
two important concepts in the theoretical description of charged
macromolecules. In this brief contribution to Ref. \cite{les_houches} 
we provide an introduction
to both ideas and summarize a few important results which can be
obtained from them. Our article is organized as follows: Section
\ref{sec:cell} outlines the sequence of approximations which ultimately
lead to the cell-model. Section \ref{sec:exactResults} is devoted to
two exact results, namely, an expression for the osmotic pressure\index{Osmotic pressure} and
a formula for the ion density at the surface of the macromolecule,
known as the contact value theorem.\index{Contact value theorem} Section \ref{sec:PB} provides a
derivation of the Poisson-Boltzmann equation from a variational
principle and the assumption of a product state.\index{Product state} Section
\ref{sec:rodPB} applies Poisson-Boltzmann theory to the cell-model of
linear polyelectrolytes. In particular, the behavior of the exact
solution in the limit of zero density is compared to the concept of
Manning condensation.\index{Manning condensation} Finally, Section \ref{sec:Donnan} shows how a
system described by a cell model can be coupled to a salt reservoir,
\ie, how the so-called Donnan equilibrium\index{Donnan equilibrium} is established.

Our main motivation is to compile in a concise form a few of the basic
concepts which form the arena for more advanced theories, treated in
other lectures of this volume. Many of the basic concepts are
discussed at greater length in a review article by Katchalsky
\cite{Kat71}, which we warmly recommend.


\section{The cell model}\label{sec:cell}


\subsection{The need for approximations}

Solutions of charged macromolecules are tremendously complicated
physical systems, and their theoretical treatment from an ``ab
initio'' point of view is surely out of question. The standard solvent
itself -- water -- already poses formidable problems.  Adding the
solute requires additional understanding of the solvent-solute
interaction, the degree of dissociation of counterions, the
conformation of the macroions, its intricate coupling with the
distribution of the counterions and many further complications. How
can one ever hope to achieve even some qualitative predictions about
such systems?

Many of the interesting features of polyelectrolytes are
ultimately a consequence of the presence of charges. One may thus
hope that a theoretical description focusing entirely on a good
treatment of the electrostatics and using crude approximations for
essentially all other problems will unveil why these systems
behave the way they do. In a first important step any quantum
mechanical effects are ignored by using a classical description. A
second simplification is to treat the solvent as a dielectric
continuum and consider explicitly only the objects having a
monopole moment. This ``dielectric approximation'' is motivated by
the long-range nature of Coulomb's law and works surprisingly well
\cite{JoWe:LesHouches}. Since a classical system of point charges
having both signs is unstable against collapse, a short-range
repulsive interaction is required, which is most commonly modeled
as a hard core. For a simple electrolyte this approximation is
called the ``restricted primitive model''.

In 1923 Debye and H\"uckel studied such a system using the
linearized Poisson-Boltzmann theory \cite{DH23}. Their treatment
accounted for the fact that ions tend to surround themselves by
ions of opposite charge, which reduces the electric field of the
central ion when viewed from a distance. While the exponentially
screened Coulomb potential is one of the most prominent results,
it must be noted that the authors computed the free
energy\index{Free energy} of the electrolyte. Its electrostatic
contribution scales as the $3/2$ power of the salt density, which
explains why a virial expansion must fail.  A good textbook
account is given in Ref.~\cite{HiMcQ}.

Though approximate, Debye-H\"uckel-theory\index{Debye-H\"uckel}
works very well for 1:1 electrolytes. However, perceptible
deviations are already much larger in the 1:2 case. This is not
merely a consequence of the presence of multivalent ions, namely
that the increased strength of electrostatic interaction may
correlate ions more strongly. Rather, the {\em asymmetry\/} of the
situation itself is a key source of the problem -- see
Kjellander's lecture for more details \cite{Kjel:LesHouches}. It
is for this very reason that in the highly asymmetric case of a
charged macromolecule surrounded by small counterions the standard
Debye-H\"uckel theory cannot be applied.


\subsection{Decoupling the macroions}

The cell model is an attempt to turn this situation into an advantage:
If the situation is highly asymmetric, there is no reason to pursue a
symmetric treatment. Since the macroions all have the same charge,
their mutual pair interaction is repulsive. Unless there are effects
which cause them to attract, aggregate and ultimately fall out of
solution, \ie, unless the effective pair potential is no longer
repulsive, the macroions will organize so as to keep themselves as far
apart as possible. The total solution can now be partitioned into
cells, each containing one macroion, the right amount of counterions
to render the cell neutral, and possibly salt molecules as well.  As a
consequence of the assumed homogeneous distribution of macroions,
these cells will all have essentially the same volume, equal to the
total volume divided by the number of macroions. Observe that
different cells do not have strong electrostatic interactions, since
they are neutral by construction.

\begin{figure}[t]
  \begin{center}
    \includegraphics[scale=0.85]{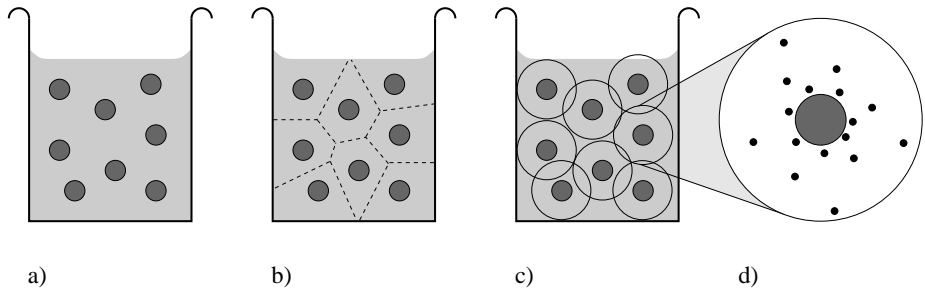}
  \end{center}
  \caption{\small Approximation stages of the cell model. The full
  solution (a) is partitioned into cells (b), which are conveniently
  symmetrized (c). Subsequently the attention is restricted to just
  one such cell (d) and the counterion distribution within it.}
  \label{fig:sph_cell}
\end{figure}

The cell model approximation consists in restricting the
theoretical description of the total system to just one cell.
While the interactions between the small ions with ``their''
macroion as well as with small ions in the same cell are
explicitly taken into account, all interactions across the cell
boundary are neglected.  Note that the existence of cells, all of
which have essentially the same size, requires correlations to be
present between the macroions. However, these correlations are no
longer the subject of study. The cell model can thus be viewed as
an approximate attempt to factorize the partition
function\index{Partition function} in the macroion coordinates,
\ie, replacing the many polyelectrolyte problem by a one
polyelectrolyte problem. Figure~\ref{fig:sph_cell} (a)
$\rightarrow$ (d) illustrates this process for the spherical case.

The remaining effect of all other macroions is to determine the
{\em volume\/} of the cell, but so far nothing has been said about
its {\em shape}. It is conveniently chosen so as to simplify
further progress. For instance, in computer simulations the cell
could be identified with the replicating unit of periodic boundary
conditions. This requires space-filling cells, for instance,
cubes. In an analytical treatment one usually tries to maximize
the symmetry of the problem. Hence, spherical colloids are
centered in spherical cells, see Fig.~\ref{fig:sph_cell} (c). The
main advantage of this strategy is that a density-functional
approach neglecting symmetry-breaking fluctuations becomes a
one-dimensional problem. Linear polyelectrolytes are enclosed in
cylindrical cells, but it is more difficult to give a precise
meaning to these cylinders -- they may not even exist in the
solution. For instance, even if the polyelectrolyte is very stiff
and thus locally straight (like, \eg, DNA), the whole molecule can
be fairly coiled on larger scales. In this case one may look at
the cylindrical tube enclosing the molecule and simply neglect the
fact that it is bent on scales larger than the persistence length
of the polyelectrolyte, provided that the tube diameter is small
compared to the latter. We thereby pretend that a locally rod-like
object is also globally straight. This point of view is justified
as long as the observables that we set out to calculate are
dominated by local, effectively short-ranged, interactions---as is
the case for ion profiles close to the macroion or the osmotic
pressure of the counterions.  Care must be exercised for
observables that may depend on the actual global shape of the
macroion, like for instance the viscosity of the solution.

In the cylindrical case a very common further approximation is to
neglect end-effects at the cylinder caps by assuming the cylinders to
be infinitely long. This additional approximation can of course only
be good if the actual finite cylinders are much longer than they are
wide. Obviously, the aim of all these approximations is to capture the
dominant effect of a locally cylindrical electrical field that an
elongated charged object generates.


\section{Some exact results}\label{sec:exactResults}

Although the partition function for systems with interacting
degrees of freedom can only be evaluated in very special cases, it
is frequently possible to derive rigorous relations which the
exact solution has to satisfy. For restricted primitive
electrolytes there exist \eg\ the Stillinger-Lovett moment
conditions \cite{StLo68}, which pose restrictions on the integral
over the ion-ion correlation functions and its second moment, or
extensions of these conditions to non-uniform electrolytes
\cite{CaCh81}.  Such results are of great theoretical interest,
since they can be used as a consistency test for approximate
theories. They can also be used to check simulations and may
provide very direct ways for analyzing them.

In this section we give the derivation of two exact results which are
particularly relevant for the cell-model.  The first is an exact
expression for the osmotic pressure in terms of the particle density
at the cell boundary.  The second is known as the ``contact value
theorem'' and provides a relation between the osmotic pressure, the
particle density, and the surface charge density at the point of
contact between the macroion and the electrolyte. It has first been
derived by Henderson and Blum \cite{HeBl78} within the framework of
integral equation theories and later by Henderson \etal\ \cite{HeBl79}
using more general statistical mechanical arguments.

Wennerstr\"om \etal\ \cite{WeJo82} give a transparent proof of both
results based on the fact that derivatives of the free energy with
respect to the cell boundaries can be expressed in terms of simple
observables. Their argument goes as follows: Let $A_R$ denote the area
of the outer cell boundary and $R$ its position, such that
infinitesimal changes $\romd R$ change the cell volume by $A_R \romd
R$.  The free energy is given by $F=-k_\romB T \, \ln Z$, where $Z =
\romTr\,\{\rome^{-\beta H}\}$ is the canonical partition function, $H$
is the Hamiltonian, $\beta\equiv 1/k_\romB T$, and $\romTr\,\{\cdot\}$
is the integral (``trace'') over phase space. The pressure is then
given by
\begin{eqnarray}
  P \; = \; - \frac{\partial F}{\partial V}
    \; = \;   \frac{k_\romB T}{A_RZ}\,\frac{\partial Z}{\partial R}.
  \label{eq:P}
\end{eqnarray}
It is easy to see that the energy of the system is independent of the
location of the outer cell boundary, since it is hard and carries no
charge. Hence, $R$ enters the partition function only via the upper
boundaries in the configuration integrals, and the derivative of $Z$
with respect to $R$ can be transformed according to
\begin{eqnarray}
  \frac{\partial Z}{\partial R}
    & = &
  \frac{\partial}{\partial R} \; \romTr\,\exp\Big\{\!\!-\beta
    H\big(\VECr_1,\ldots,\VECr_N\big)\Big\}
  \nonumber \\
    & = &
  A_R \; \sum_{i=1}^N \, \romTr_{\text{\tiny not $\VECr_i$}} \exp\Big\{\!\!-\beta
    H\big(\VECr_1,\ldots,\VECr_i\rightarrow R,\ldots,\VECr_N\big)\Big\}
  \nonumber \\
    & = &
  A_R \, N \; \romTr_{\text{\tiny not $\VECr_1$}} \exp\Big\{\!\!-\beta
    H\big(\VECr_1\rightarrow R,\VECr_2,\ldots,\VECr_N\big)\Big\}
  \nonumber \\[0.5em]
    & = &
  A_R \, Z \; n(R).
  \label{eq:dZdR}
\end{eqnarray}
The trace over phase space is an $N$-fold volume integral over all
particle coordinates, each containing a radial integration from $r_0$
to $R$. As a consequence of the product rule, the derivative with
respect to $R$ is the sum of $N$ terms, in which the integral from
$r_0$ to $R$ over the radial coordinate of particle $i$ is
differentiated with respect to $R$, \ie, the integration is omitted
and the radial coordinate is set to $R$. The integration over the two
remaining coordinates now yield a prefactor $A_R$.  Since all
particles are identical, these $N$ terms are all equal. In the last
step we used the fact that the trace over all particles but the first
one is equal to $Z$ times the probability distribution of the first
particle, so a multiplication by $N$ gives the ion density $n(R)$ at
the cell boundary. Combining Eqns.~(\ref{eq:P}) and (\ref{eq:dZdR}) we
obtain the pressure:
\begin{eqnarray}
  \beta P \; = \; n(R).
  \label{eq:PR}
\end{eqnarray}
In words: The osmotic pressure in the cell-model is exactly given
by $k_\romB T$ times the particle density at the outer cell
boundary.  If more than one species of particles are present,
$n(R)$ is replaced by the sum $\sum_i n_i(R)$ over the boundary
densities of these species. Note that despite its ``suggestive''
form Eqn.~(\ref{eq:PR}) does by no means state that the particles
at the outer cell boundary behave like an ideal gas. Even if the
system is dense and the particles are strongly correlated,
Eqn.~(\ref{eq:PR}) is valid, since it is completely independent of
the pair interactions entering $H$.

In a similar fashion one can compute the derivative of the free energy
with respect to the inner cell boundary, \ie, the location of the
surface of the macroion. In this case the geometry enters the problem,
since in the non-planar case a change in the location of this surface
necessarily also changes the surface charge density, if the total
charge is to remain the same. For simplicity we will restrict
ourselves to the planar case here and defer the reader to
Ref.~\cite{WeJo82} for the other geometries.

Let $A_{r_0}$ be the area of the inner cell boundary and $r_0$ the
position of the surface of the macroion, such that an infinitesimal
positive change $\romd r_0$ makes the macroion larger, but reduces the
volume available for the counterions by $A_{r_0} \romd r_0$.  The
pressure is thus given by
\begin{eqnarray}
  P \; = \; - \frac{\partial F}{\partial V}
    \; = \; - \frac{k_\romB T}{A_{r_0}Z}\,\frac{\partial Z}{\partial r_0}.
  \label{eq:Pinner}
\end{eqnarray}
The key difference with the previous case is that the energy of an ion
also depends on the location of the wall, since the latter is
charged. Hence, our calculation leading to Eqn.~(\ref{eq:dZdR}) must
be supplemented by an additional term $\romTr\big[\frac{\partial}{\partial r_0}
\rome^{-\beta H}\big]$, which leads to the expression
\begin{eqnarray}
  \frac{\partial Z}{\partial r_0} \; = \; - A_{r_0} \, Z \; n(r_0) \;
  - \; \frac{Z}{k_\romB T}\left\langle\frac{\partial H}{\partial
  r_0}\right\rangle.  \label{eq:dZdr0}
\end{eqnarray}
It is easy to see that $\partial H/\partial r_0 = -2 \pi \ell_\romB
\tilde{\sigma}^2 A_{r_0}$, independent of the ion coordinates. Here,
$\ell_\romB=\beta e^2/4\pi\varepsilon_0\varepsilon_\romr$ is the
Bjerrum length, \ie, the distance at which two unit charges have
interaction energy $k_\romB T$, and $\tilde{\sigma}$ is the number
density of surface charges. Combining this with
Eqns.~(\ref{eq:Pinner}) and (\ref{eq:dZdr0}) finally gives
\begin{eqnarray}
  \beta P \; = \; n(r_0) - 2 \pi \ell_\romB \tilde{\sigma}^2.
  \label{eq:Pr0}
\end{eqnarray}
This equation is known as the contact value theorem, since it gives
the contact density at a planar charged wall as a function of its
surface charge density and the osmotic pressure. The occurrence of the
second term is related to the presence of an electric field, which
vanishes at the outer cell boundary and which contributes its share to
the total pressure via the Maxwell stress tensor
\cite{LaLiel}. Observe finally that by subtracting
Eqns.~(\ref{eq:PR}) and (\ref{eq:Pr0}) we obtain a relation between
the ion density at the inner and outer cell boundary. Taking into
account different ion species, it reads
\begin{eqnarray}
  \sum_i n_i(r_0) - \sum_i n_i(R) \; = \; 2 \pi \ell_\romB \tilde{\sigma}^2.
  \label{eq:Graham}
\end{eqnarray}
This is a rigorous version of an equation which has been derived on
the level of Poisson-Boltzmann theory by Grahame \cite{Gra47}.  Note
that since the densities $n_i(R)$ are bounded below by $0$, the
contact density is at least $2\pi\ell_\romB\tilde{\sigma}^2$.

We would like to emphasize that Eqns.~(\ref{eq:Pr0}) and
(\ref{eq:Graham}) only apply to the planar case. For a cylindrical or
spherical geometry the contact density for the same values of
$\tilde{\sigma}$ and $P$ is lower \cite{WeJo82}. We will briefly
return to this point in Sec.~\ref{sec:PB_limits}.


\section{Poisson-Boltzmann theory}\label{sec:PB}

What makes the computation of the partition function so extremely
difficult? It is the fact that all ions interact with each other,
implying that their positions are mutually correlated. Stated
differently, the many-particle probability distribution does not
factorize into single-particle distributions, and hence the partition
function does not factorize in the ion coordinates. Poisson-Boltzmann
theory is the mean-field route to circumventing this problem. Its
following derivation demonstrates this point in a particularly clear
way. We largely follow the lines of Ref.~\cite[Ch.~4.8]{ChLu95}.

Quite generally, the free energy $F$ can be bounded from above by
\cite{GB}:
\begin{eqnarray}
  F \; \le \; \langle H \rangle_0 - T \, S_0,
  \label{eq:GibbsBogoliubov}
\end{eqnarray}
where $\langle H \rangle_0 = \romTr\,\{p_0 H\}$ is the expectation
value of the energy in some arbitrary state existing with
probability $p_0$ and $S_0 = -k_\romB \, \romTr \, \{p_0\ln p_0\}$
is the entropy of that state. This relation is sometimes referred
to as the Gibbs-Bo\-go\-liu\-bov-in\-equa\-li\-ty and provides a
general and powerful way of deriving mean-field theories from a
variational principle \cite{ChLu95}. Its equality version holds if
and only if $p_0$ is the canonical probability $\rome^{-\beta
H}\!/\,\romTr\,\{\rome^{-\beta H}\}$.

Assume we have a system of $N$ point-particles of charge $ze$ and mass
$m$ within a volume $V$, and additionally some fixed charge density $e
n_\romf(\VECr)$.  The Hamiltonian $H$ is the sum of the kinetic energy
$K(\VECp_1,\ldots,\VECp_N)$ and the potential energy
$U(\VECr_1,\ldots,\VECr_N)$, and up to an irrelevant additive constant
it is given by
\begin{eqnarray}
H & = & \sum_{i=1}^N \frac{\VECp_i^2}{2m} \; + \;
 \sum_{i<j=1}^N\frac{z^2e^2}{4 \pi \varepsilon_0\varepsilon_\romr
   |\VECr_i-\VECr_j|} \; + \;
 \int_V \romd^3 r \sum_{i=1}^N \frac{z n_\romf(\VECr)e^2}{4 \pi
   \varepsilon_0\varepsilon_\romr |\VECr_i-\VECr|} \nonumber \\
{} & = & \sum_{i=1}^N \frac{\VECp_i^2}{2m} \; + \;
 z e \sum_{i=1}^N \Big(\frac{1}{2}\,\psi(\VECr_i)+\psi_\romf(\VECr_i)\Big),
\end{eqnarray}
where $\psi$ and $\psi_\romf$ are the electrostatic potentials
originating from the ions and from the fixed charge density,
respectively. In a classical description position and momentum are
commuting observables, so the momentum part of the canonical partition
function factorizes out. Since this is just a product of $N$ identical
Gaussian integrals, its contribution to the free energy is readily
found to be
\begin{eqnarray}
  \beta F_\VECp
    \; = \;
  - \ln \, \romTr_\VECp\,\big[\rome^{-\beta K}\big]
    \; = \;
  \ln(N!\,\lambda_\romT^{3N})
    \; \simeq \;
  N \, \Big[ \ln (N \lambda_\romT^3) - 1 \Big]
  \label{eq:Fp}
\end{eqnarray}
where $\lambda_\romT = h/\sqrt{2\pi m k_\romB T}$ is the thermal
de\,Broglie wavelength, and where Stirling's approximation $\ln N!
\simeq N\,\ln N - N$ has been used in the last step.

The complication comes from the $\VECr$-part of the partition
function, specifically from the fact that the couplings between the
positions $\VECr_i$ appearing in $U$ render the $N$-particle
distribution function $p_N(\VECr_1,\ldots,\VECr_N)\equiv \rome^{-\beta
U}\!/\,\romTr_\VECr \, \big[\rome^{-\beta U}\big]$ essentially
intractable. The purpose of any mean-field approximation is to remove
these correlations between the particles. One way of achieving this
goal is by replacing the $N$-particle distribution function by a
product of $N$ identical one-particle distribution functions:
\begin{eqnarray}
  p_N(\VECr_1,\ldots,\VECr_N)
  \;\; \stackrel{\text{\footnotesize mean-field}}{\longrightarrow} \;\;
  p_1(\VECr_1) \, p_1(\VECr_2) \, \cdots \, p_1(\VECr_N).
  \label{eq:meanfield}
\end{eqnarray}
Of course, this {\em product state\/} is different from the canonical
state, but if used as a trial state in the
Gibbs-Bo\-go\-liu\-bov-in\-equa\-li\-ty (\ref{eq:GibbsBogoliubov}) it
yields an upper bound for the free energy. The electrostatic
contribution $\langle U \rangle_0$ is then given by
\begin{eqnarray}
  \langle U \rangle_{0} & = &
  \int_V \romd^3 r_1 \cdots \int_V \romd^3 r_N \;
  p_1(\VECr_1) \, \cdots \, p_1(\VECr_N) \;
  z e \sum_{i=1}^N \Big(\frac{1}{2}\,\psi(\VECr_i)+\psi_\romf(\VECr_i)\Big)
  \nonumber \\ & = &
  N \; \int_V \romd^3 r \; p_1(\VECr) \; z e \;
  \Big(\frac{1}{2}\,\psi(\VECr)+\psi_\romf(\VECr)\Big).
  \label{eq:H0}
\end{eqnarray}
Similarly, the entropy $S_0$ is found to be
\begin{eqnarray}
  S_0 \;\;\; & = &
  -k_\romB \; \int_V \romd^3 r_1 \cdots \int_V \romd^3 r_N \;
  p_1(\VECr_1) \, \cdots \, p_1(\VECr_N) \;
  \ln\big(p_1(\VECr_1) \, \cdots \, p_1(\VECr_N)\big)
  \nonumber \\ & = &
  - N \, k_\romB \; \int_V \romd^3 r \; p_1(\VECr)\ln\big(p_1(\VECr)\big).
  \label{eq:S0}
\end{eqnarray}
Notice the key effect of the factorization assumption
(\ref{eq:meanfield}): It reduces an $N$-dimensional integral to $N$
identical one-dimensional integrals, thereby making the problem
tractable. Observe also that by definition the one-particle
distribution function is proportional to the density:
\begin{eqnarray}
  p_1(\VECr) \; \equiv \;
  \frac{n(\VECr)}{\int_V \romd^3 r \; n(\VECr)} \; = \;
  \frac{n(\VECr)}{N}
  \label{eq:w1}
\end{eqnarray}
Combining Eqns.~(\ref{eq:GibbsBogoliubov}), (\ref{eq:Fp}),
(\ref{eq:H0}), (\ref{eq:S0}) and (\ref{eq:w1}), we arrive at the
following bound for the free energy:
\begin{eqnarray}
  F \; \le \; F_\romPB[n(\VECr)],
  \label{eq:Fn_ineq}
\end{eqnarray}
where the Poisson-Boltzmann density functional\index{Density functional} is given by
\begin{eqnarray}
  F_\romPB[n(\VECr)]
  & = &
  \int_V \romd^3 r \; \bigg\{
    z e n(\VECr)\Big[\frac{1}{2}\,\psi(\VECr)+\psi_\romf(\VECr)\Big]
  \nonumber \\
  & & \qquad\;\; + \;\;
    k_\romB T \, n(\VECr) \,
    \Big[ \ln \big(n(\VECr)\lambda_\romT^3\big)-1\Big]\bigg\}.
  \label{eq:Fn}
\end{eqnarray}
Clearly, we aim for the best -- \ie\ lowest -- upper bound; we
therefore want to know which density $n(\VECr)$ minimizes this
functional. Hence, the mean-field approach has led us to the
variational problem of {\em minimization of a density
functional}. Setting the functional derivative $\delta F_\romPB[n] /
\delta n$ to zero and requiring that ($i$) the charge density and the
electrostatic potential be related by Poisson's equation (see below)
and ($ii$) the total number of particles be $N$ leads finally to the
Poisson-Boltzmann equation.\footnote{Note the mathematical analogy in
classical mechanics, where the Euler-Lagrange differential equations
correspond to Hamilton's variational principle of a stationary action
functional.}

In order to fulfill the first constraint, we add a term $\mu_0 \,
(n(\VECr)-N/V)$ to the integrand in Eqn.~(\ref{eq:Fn}), where $\mu_0$ is a
Lagrange multiplier. The second constraint is automatically satisfied
if we rewrite $\psi(\VECr)$ in terms of $n(\VECr)$. The functional
derivative then gives:
\begin{eqnarray}
  0
  \; = \;
  \frac{\delta F[n(\VECr)]}{\delta n(\VECr)}
  \; = \;
  \mu(\VECr)
  \; = \;
  \mu_0 \, + \,
  z e  \, \psi_\romtot(\VECr) \; + \;
  k_\romB T \, \ln \big(n(\VECr)\lambda_\romT^3\big),
  \label{eq:mur}
\end{eqnarray}
where $\psi_\romtot=\psi+\psi_\romf$ is the total electrostatic
potential. We could have ``guessed'' this equation right away from
a close inspection of the free energy density in
Eqn.~(\ref{eq:Fn}), which consists of only two simple terms: The
first is the electrostatic energy of a charge distribution $e
n(\VECr)$ in the potential created by itself and by an additional
external potential $\psi_\romf(\VECr)$; the second is the entropy
of an ideal gas with density $n(\VECr)$.  Stated differently,
apart from the fact that the particles are charged, we are
effectively dealing with an ideal gas. In fact, the right hand
side of Eqn.~(\ref{eq:mur}) is just the (local) electrochemical
potential of a system of charged particles in the ``ideal gas
approximation'', \ie, assuming that the activity coefficient is
equal to $1$.

The condition $\mu(\VECr)=0$ can be written in a more familiar way:
\begin{eqnarray}
  n(\VECr)
    \; = \;
  \lambda_\romT^{-3} \rome^{-\beta ( z e \, \psi_\romtot(\VECr) + \mu_0)}
    \; = \;
  n_0\,\rome^{-\beta z e \, \psi_\romtot(\VECr)}
  \label{eq:Boltzmann}
\end{eqnarray}
where $\mu_0$ or $n_0$ are fixed by the equation $\int \romd^3 r \;
n(\VECr) = N$, \ie, by the requirement of particle conservation.  Note
that $n_0$ is the particle density at a point where $\psi_\romtot =
0$.  Eqn.~(\ref{eq:Boltzmann}) states that the ionic density is
locally proportional to the Boltzmann factor. Although this appears to
be a very natural equation, which is taken for granted in most
``derivations'' of the Poisson-Boltzmann equation, the above
derivation shows it to be the result of a mean-field treatment of the
partition function.

Combining Eqn.~(\ref{eq:Boltzmann}) with Poisson's equation $\Delta
\psi_\romtot(\VECr) = - e \big(z n(\VECr)+n_\romf(\VECr)\big) /
\varepsilon_0\varepsilon_\romr$ yields the Poisson-Boltzmann equation
\begin{eqnarray}
  \Delta \psi_\romtot(\VECr) \; = \; -
  \frac{e}{\varepsilon_0\varepsilon_\romr} \Big[ \; z n_0 \,
  \rome^{-\beta z e \, \psi_\romtot(\VECr)} \, + \, n_\romf(\VECr) \;
  \Big].  \label{eq:PB}
\end{eqnarray}
The standard situation is that the counterions are localized within
some region of space, outside of which there is a fixed charge
distribution. In this case one solves Eqn.~(\ref{eq:PB}) within the
inner region, where $n_\romf \equiv 0$, and incorporates the effects
of the outer charges as a boundary condition to the differential
equation.

\vspace*{0.5em}

We would finally like to show that Eqn.~(\ref{eq:PR}), the rigorous
expression for the osmotic pressure of the cell-model in terms of the
boundary density, is also valid on the level of Poisson-Boltzmann
theory. We therefore have to compute the derivative of the
Poisson-Boltzmann free energy with respect to the volume. Since we
again assume the outer cell boundary to be hard and neutral, the
energy of the system (in particular: $\psi_\romf(\VECr)$) does not
explicitly depend on its location $R$, which hence only enters the
boundaries in the volume integration. But changing the volume of the
cell could entail a redistribution of the ions, which also may change
the free energy. Let us symbolically express this in the following
way:
\begin{eqnarray}
  \delta F
    \; = \;
  \frac{\partial F}{\partial V}\,\delta V
    \; + \;
  \int \romd^3 r \; \frac{\delta F}{\delta n(\VECr)}\,\delta n(\VECr).
\end{eqnarray}
However, since the Poisson-Boltzmann profile renders the functional
$F$ stationary, the second contribution vanishes. The pressure is thus
given by
\begin{eqnarray}
  P \; = \;
  - \frac{1}{A_R}\frac{\partial F}{\partial R}\bigg|_{\text{\tiny PB-profile}},
  \label{eq:PB_pressure}
\end{eqnarray}
\ie, the free energy functional is differentiated with respect to the
outer cell boundary and evaluated with the Poisson-Boltzmann profile.
The quantity $A_R$ is the area of the outer boundary, as introduced in
Sec.~\ref{sec:exactResults}. If one again rewrites $\psi(\VECr)$ in
terms of $n(\VECr)$, the derivative is readily found to be
\begin{eqnarray}
  \frac{1}{A_R}\frac{\partial F}{\partial R}
    \; = \;
 z e n(R) \, \psi_\romtot(R) + k_\romB T \,
  n(R) \, \Big[\ln\big(n(R)\lambda_\romT^3\big)-1\Big] + \mu_0\,n(R).
\end{eqnarray}
Together with the Poisson-Boltzmann profile from
Eqn.~(\ref{eq:Boltzmann}) and Eqn.~(\ref{eq:PB_pressure}) this finally
yields $P=n(R)k_\romB T$, as we had set out to show.

The first derivation of this result was given by Marcus
\cite{Mar55}. Its intuitive interpretation is as follows: The
Poisson-Boltzmann free energy functional describes a system of
charged particles in the ideal gas approximation.  Since the
pressure is constant, we may evaluate it everywhere, \eg\ at the
outer cell boundary, where the electric field vanishes. The latter
implies that the rod exerts no force on the ions sitting there, so
the only remaining contribution to their pressure is the ideal gas
equation of state evaluated at the local density $n(R)$.

It is by no means trivial that Poisson-Boltzmann theory gives the same
relation between boundary density and pressure. Rather, it is one of
its pleasant features that it retains this exact result.  This does of
course not mean that Poisson-Boltzmann theory gives the correct
osmotic pressure, since its prediction of the boundary density is not
correct.

\vspace*{0.5em}

The Poisson-Boltzmann equation has been and remains extremely
important as a mean-field approach to charged systems. The above
derivation shows how it neglects all correlations (see
Eqn.~(\ref{eq:meanfield})) and that their incorporation will decrease
the free energy (see Eqn.~(\ref{eq:Fn})). In the lecture notes of
Moreira and Netz \cite{MoNe:LesHouches} a different derivation is
presented, which shows the Poisson-Boltzmann theory to be the
saddle-point approximation of the corresponding field-theoretic
action. This latter approach nicely clarifies those observables that
have to be small in order for Poisson-Boltzmann theory to be a good
approximation --- and also what to do if these parameters happen to be
large.


\section{Concrete example: Poisson-Boltzmann theory for charged\break rods}\label{sec:rodPB}

In this section we will apply the cell model and its Poisson-Boltzmann
solution to the case of linear polyelectrolytes. The main purpose is
to demonstrate how the theoretical considerations presented so far can
be applied to a realistic situation. We chose the cylindrical geometry
since this gives us the opportunity to compare Poisson-Boltzmann
theory with Manning's scenario of counterion condensation.  A more
comprehensive and very readable introduction, also covering other
geometries, is given in Ref.~\cite{And95}.


\subsection{Specification of the cell model}

Consider linear polyelectrolytes of line charge density $\lambda=e
\tilde{\lambda} > 0$, radius $r_0$, and length $L$ that are distributed
at a density $n_\romP$ in a solvent characterized by a Bjerrum length
$\ell_\romB$. We will assume the counterions to be monovalent
($z=-1$), and at first the system does not contain additional
salt. Let us enclose the polyelectrolytes by cylindrical cells of
radius $R$ and length $L$. Requiring that the volume of each cell
equals the volume per polyelectrolyte in the original solution gives
the relation $n_\romP = 1/\pi R^2 L$, from which we derive the cell
radius.  If the polyelectrolytes are bent on a large scale, we will
require the cell radius $R$ to be small compared to the persistence
length of the charged chains and subsequently neglect the bending. We
will thus refer to the linear polyelectrolytes simply as ``charged
rods''.

\begin{figure}[t]
  \begin{center}
    \includegraphics[scale=0.85]{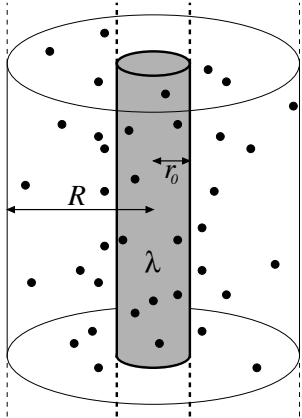}
  \end{center}
  \caption{\small Geometry of the cell model. A cylindrical rod with
  radius $r_0$ and line charge density $\lambda$ is enclosed by a
  cylindrical cell of radius $R$. Note that within Poisson-Boltzmann
  theory the ions are point-like in the sense that no
  hard core energy term enters the free energy functional $F_\romPB$
  from Eqn.~(\ref{eq:Fn}). However, the theory does not describe
  individual point-ions but rather an average ionic density.}\label{fig:cyl_cell}
\end{figure}

For monovalent counterions each of these rods dissociates
$N_\romc=\tilde{\lambda} L$ of them into the solution, such that their
average density is given by $\bar{n}_\romc=N_\romc n_\romP =
\tilde{\lambda} / \pi R^2$. If one now neglects end-effects, \ie, if
one sets $L$ to infinity, the problem acquires cylindrical
symmetry. Observe that $\bar{n}_\romc$ is independent of $L$. It is
therefore a more convenient measure of the system density, since it is
unaffected by the limit $L\rightarrow\infty$.

The next step is to replace the individual ion coordinates by a
density $n(\VECr)$. The cylindrical symmetry will be exploited by
assuming that this density also has cylindrical symmetry, even
further, that it only depends on the radial coordinate. It is
worthwhile pointing out that this statement is not a trivial
assumption, for two reasons: First, just because the problem is
cylindrically symmetric, the solution need not
be.\footnote{Broadly speaking: If a physical problem is invariant
with respect to some symmetry group $\cal{S}$, any solution of the
problem is mapped by any element of $\cal{S}$ to another
solution---the set of solutions is closed under $\cal{S}$.
However, a particular solution need not be mapped onto itself by
{\em all\/} elements of $\cal{S}$, \ie, it need not possess the
full symmetry of the problem.  Let us give a simple example: The
gravitational field of our sun is spherically symmetric, but the
orbit of the earth is not (even if time-averaged).}  Second, even
if the average distribution only depends on the radial coordinate,
there may be angular or axial fluctuations that are not taken into
account if one right from the start only works with one radial
coordinate.


\subsection{Solution of the Poisson-Boltzmann equation}

Employing these approximations, the Poisson-Boltzmann equation
(\ref{eq:PB}) in the region between $r_0$ and $R$ can be written as
\begin{eqnarray}
  y''(r) + \frac{1}{r} \, y'(r) \; = \; \kappa^2 \rome^{y(r)}.
  \label{eq:PB_cyl}
\end{eqnarray}
Here, $\kappa=\sqrt{4\pi\ell_\romB n(R)}$ is an inverse length (the
Debye screening constant at the outer boundary) and $y(r)=\beta e
\psi(r)$ is the dimensionless potential, which is understood to be
zero at $r=R$. It will turn out to be convenient to introduce the
dimensionless charge parameter
\begin{eqnarray}
  \xi \; = \; \tilde{\lambda}{\ell_\romB}.
\end{eqnarray}
It counts the number of charges along a Bjerrum length of rod and is
thus a dimensionless way to measure the line charge density $\lambda$.

The boundary conditions at $r=r_0$ and $r=R$ follow easily, since by
Gauss' theorem we know the value of the electric field there:
\begin{eqnarray}
  y'(r_0) \; = \; -\frac{2\xi}{r_0}
  \qquad\quad\text{and}\quad\qquad
  y'(R) \; = \; 0.
  \label{eq:PB_cyl_boun}
\end{eqnarray}

The nonlinear boundary value problem that Eqns.~(\ref{eq:PB_cyl}) and
(\ref{eq:PB_cyl_boun}) pose was first solved independently by Fuoss
\etal\ \cite{FuKa51} and Alfrey \etal\ \cite{AlBe51}. It can easily be
verified by insertion that the solution is given by
\begin{eqnarray}
  y(r) \; = \; -2 \, \ln \left\{\frac{r}{R}\sqrt{1+\gamma^{-2}}\,
    \cos\Big(\gamma\,\ln\frac{r}{R_\romM}\Big) \right\}.
  \label{eq:PB_sol}
\end{eqnarray}
The dimensionless integration constant $\gamma$ is related to $\kappa$ via
\begin{eqnarray}
  \kappa^2R^2 \; = \; 2\,(1+\gamma^2).
  \label{eq:kappagamma}
\end{eqnarray}
Both $\gamma$ and $R_\romM$ are found by inserting the general
solution (\ref{eq:PB_sol}) into the boundary conditions
(\ref{eq:PB_cyl_boun}), which yields two coupled transcendental
equations:
\begin{eqnarray}
  \gamma\,\ln\frac{r_0}{R_\romM} \; = \; \arctan\frac{1-\xi}{\gamma}
  \qquad\text{and}\qquad
  \gamma\,\ln\frac{R}{R_\romM} \; = \; \arctan\frac{1}{\gamma}.
  \label{eq:PB_boun_eq}
\end{eqnarray}
Subtracting them eliminates $R_\romM$ and provides a single equation
from which $\gamma$ can be obtained numerically:
\begin{eqnarray}
  \gamma\,\ln\frac{R}{r_0} \; = \;
  \arctan\frac{1}{\gamma} + \arctan\frac{\xi-1}{\gamma}.
  \label{eq:PB_gamma}
\end{eqnarray}

Eqn.~(\ref{eq:PB_gamma}) only has a real solution for $\gamma$ if $\xi
> \xi_{\min} = \ln(R/r_0)/(1+\ln(R/r_0))$. For smaller charge
densities $\gamma$ becomes imaginary, but the solution can be
analytically continued by replacing $\gamma \rightarrow \romi\gamma$
and using identities like $\romi\gamma\,\tan(\romi\gamma) =
-\gamma\,\tanh\gamma$. However, in the following we will only be
interested in the strongly charged case, in which the charge parameter
$\xi$ is larger than $\xi_{\min}$.

Let us denote by $\phi(r)$ the fraction of counterions that can be
found between $r_0$ and $r$. Using $n(r)=n(R)\exp\{y(r)\}$ and
Eqns.~(\ref{eq:PB_sol}), (\ref{eq:kappagamma}), and
(\ref{eq:PB_boun_eq}), we find
\begin{eqnarray}
  \phi(r)
    \; = \;
  \frac{1}{\tilde{\lambda}}\int_{r_0}^r \romd\bar{r}\;2\pi\bar{r}\,n(\bar{r})
    \; = \;
  1 - \frac{1}{\xi} + \frac{\gamma}{\xi}\tan\Big(\gamma\,\ln\frac{r}{R_\romM}\Big).
  \label{eq:phi(r)}
\end{eqnarray}
Observe that $\phi(R_\romM)=1-1/\xi$. Hence, the second integration
constant $R_\romM$ is the distance at which the fraction $1-1/\xi$ of
counterions can be found, which also implies $r_0 \le R_\romM < R$.
Due to the importance of this fraction in Manning's theory of
counterion condensation (see Sec.~(\ref{ssec:Manning})), $R_\romM$ is
sometimes referred to as the ``Manning radius''.


\subsection{Manning condensation}\label{ssec:Manning}\index{Manning condensation}

The ion distribution around a charged cylindrical rod exhibits a
remarkable feature that can be unveiled by the following simple
considerations \cite{RolMan}. Assume that the system is infinitely
dilute and that there is only one counterion. In the canonical
ensemble its radial distribution should be given by $\rome^{-\beta
H(r)}/\,\romTr \, \big[\rome^{-\beta H(r)}\big]$ where, up to the
kinetic energy and an additive constant, the Hamiltonian is $\beta
H(r) = 2\xi\,\ln(r/r_0)$. However, the trace (per unit length)
over the coordinate space is
\begin{eqnarray}
  \romTr\,\big[\rome^{-\beta H}\big]
  \; = \;
  \int_{r_0}^\infty \romd r \; 2\pi r \, \rome^{-2\xi\,\ln(r/r_0)}
  \; = \;
  2\pi r_0^2 \int_1^\infty \romd x \; x^{1-2\xi},
\end{eqnarray}
which diverges for $\xi < 1$. Hence, the distribution function cannot
be normalized. In other words, such rods cannot localize counterions
in the limit of infinite dilution, while rods with $\xi>1$ can. This
led Manning to the simple idea that rods with $\xi>1$ ``condense'' a
fraction of $1-1/\xi$ of all counterions, thereby reducing
(``renormalizing'') their charge parameter to an effective value of
$1$, while the rest of the ions remains more or less ``free'', \ie,
not localized \cite{Man69}.  This concept has subsequently been
referred to as ``Manning condensation'' and has led to much insight
into the physical chemistry of charged cylindrical macroions.

Similar arguments can be made for the infinite dilution limit in the
pla\-nar and spherical case. They show that a plane always localizes
all its coun\-ter\-ions no matter how low its surface charge density
is, while a sphere always loses all its counterions no matter how high
its surface charge density is.


\subsection{Limiting laws of the cylindrical PB-solution}\label{sec:PB_limits}

Although the PB-equation for the cylindrical geometry can be solved
ana\-ly\-ti\-cal\-ly, the transcendental equation (\ref{eq:PB_gamma})
for the integration constant $\gamma$ has to be solved
numerically. However, since for $\xi>1$ its right hand side is bounded
above by its zeroth and bounded below by its first order Taylor
expansion, this gives an allowed interval for $\gamma$. All following
considerations are restricted to the strongly charged case $\xi > 1$.
\begin{eqnarray}
  \pi
    \; \ge \;
  \gamma\,\ln\frac{R}{r_0}
    & = &
  \arctan\frac{1}{\gamma} + \arctan\frac{\xi-1}{\gamma}
    \; \ge \;
  \pi - \frac{\xi}{\xi - 1}\gamma
  \qquad(\xi>1)
  \nonumber \\[1em]
  & \Rightarrow &
  \frac{\pi}{\ln\frac{R}{r_0}}
    \; \ge \;
  \gamma
    \; \ge \;
  \frac{\pi}{\ln\frac{R}{r_0}+\frac{\xi}{\xi-1}}
  \label{eq:gamma_ineq}
\end{eqnarray}
In the limit $R\rightarrow\infty$ the two bounds, and therefore
$\gamma$, converge to zero. In this limit $\gamma$ can be
approximated by either side of inequality (\ref{eq:gamma_ineq}),
which gives rise to various asymptotic behaviors, known as
``limiting laws'' for infinite dilution. An immediate first
consequence of (\ref{eq:gamma_ineq}) is that these asymptotic
behaviors are reached logarithmically slowly. In the following we
will briefly discuss four of these limiting laws.

As we have seen above, the radius $R_\romM$ contains the fraction
$1-1/\xi$ of ions that are condensed in the sense of Manning.  The
following limit shows that $R_\romM$ scales asymptotically as the
square root of the cell radius:
\begin{eqnarray}
  \lim_{R\rightarrow\infty} \frac{R_\romM}{\sqrt{R r_0}}
   \; = \;
   \exp\Big\{\frac{\xi-2}{2\xi-2}\Big\},
  \label{eq:PB_limRM}
\end{eqnarray}
It can be shown \cite{LeZi84} that a radius that is required to
contain any fraction smaller than $1-1/\xi$ will remain finite in the
limit $R\rightarrow\infty$, while a radius containing more than this
fraction will diverge asymptotically like $R$. Roughly speaking, the
fraction $1-1/\xi$ cannot be diluted away, which is in accordance with
the localization argument given in Section~\ref{ssec:Manning}.

Up to a logarithmic\footnote{Since the potential itself is already
logarithmic, the logarithmic correction is actually of the form
$\ln\,\ln r$.} correction the electrostatic potential is that of a rod
with charge parameter $1$:
\begin{eqnarray}
  \lim_{R\rightarrow\infty} y(r)
  \; = \;
  y(r_0) - 2\,\ln\frac{r}{r_0} - 2 \, \ln\Big(1 + (\xi-1)\,\ln\frac{r}{r_0}\Big).
  \label{eq:PB_limy}
\end{eqnarray}
This is Manning condensation rediscovered on the level of the
mean-field potential. Note, however, that the presence of the
logarithmic corrections implies that the condensed ions do not sit on
top of the charged rod, but rather have a radial distribution. For
finite cell radii this distribution is characterized by the length
$R_\romM$, which diverges in the dilute limit. Hence, the ions are not
particularly closely confined.

The ratio between the boundary density $n(R)$ and the average
counterion density $\bar{n}_\romc$ shows the limiting behavior
\begin{eqnarray}
  \lim_{R\rightarrow\infty} \frac{n(R)}{\bar{n}_\romc}
  \; = \;
  \lim_{R\rightarrow\infty} \frac{1+\gamma^2}{2\xi}
  \; = \; \frac{1}{2\xi}.
  \label{eq:PB_limnR}
\end{eqnarray}
Since we have seen in Sec.~\ref{sec:PB} that the boundary density is
proportional to the osmotic pressure, the ratio $n(R)/\bar{n}_\romc$
is equal to the ratio $P/P_\romid$ between the actual osmotic pressure
and the ideal gas pressure of a fictitious system of non-interacting
particles at the same average density. This ratio is called the
``osmotic coefficient'', and in the dilute limit it converges from
above towards $1/2\xi<1$. The presence of the charged rod hence
strongly reduces the osmotic activity of the counterions.

While Eqn.~(\ref{eq:PB_limnR}) implies that the boundary density goes
to zero in the dilute limit, the contact density approaches a finite
value:
\begin{eqnarray}
  \lim_{R\rightarrow\infty} n(r_0)
  \; = \;
  2\pi\ell_\romB\tilde{\sigma}^2 \; \Big(1-\frac{1}{\xi}\Big)^2
  \; = \; 2\pi\ell_\romB\tilde{\sigma}^2 \; \Big(1-
   \frac{1}{2\pi r_0\ell_\romB\tilde{\sigma}}\Big)^2.
  \label{eq:PB_limnr0}
\end{eqnarray}
This as well is a sign that ions must be condensed. We would like to
link these two equations to the contact value theorem derived in
Section~\ref{sec:exactResults}, in particular to
Eqn.~(\ref{eq:Graham}). Observe that this equation is {\em not\/}
satisfied. This is not a bug of the Poisson-Boltzmann approximation
but rather a feature of the cylindrical geometry: The contact density
is lower than in the corresponding planar case, essentially since only
the fraction $1-1/\xi$ of condensed ions ``contributes'' to the ion
distribution function. However, in the limit $r_0\rightarrow\infty$ at
{\em fixed\/} surface charge density $e\tilde{\sigma}$ the charged rod
becomes a charged plane, $\xi=2\pi\ell_\romB\tilde{\sigma}r_0$
diverges, and the correction factor becomes $1$, such that the contact
value theorem is again satisfied. Note also that $\xi$ can be written
as $r_0/\lambda_\romGC$, where $\lambda_\romGC$ is the so called
Gouy-Chapman length, the characteristic width of a planar electrical
double layer forming at a surface with surface charge density
$e\tilde{\sigma}$ \cite{And95}. For the contact value theorem to be
valid in its planar version it is thus necessary that the
characteristic extension of the charged layer is small compared to the
radius of curvature of the surface---or, equivalently: $\xi \gg 1$.
It is worth pointing out that the solution of the linearized planar
Poisson-Boltzmann equation violates the contact value theorem by
giving a contact density which is a factor of $2$ too large.


\section{Additional salt: The Donnan equilibrium}\label{sec:Donnan}\index{Donnan equilibrium}

How is the Poisson-Boltzmann equation to be modified, if more than one
species of ions is present?  First, each ion density $n_i(\VECr)$ is
assumed to be proportional to the local Boltzmann-factor, thereby
generalizing Eqn.~(\ref{eq:Boltzmann}):
\begin{eqnarray}
  n_i(\VECr) = n_{0i}\,\rome^{-\beta z_i e \, \psi_\romtot(\VECr)}
  \qquad\text{with}\qquad
  n_{0i} = \frac{N_i}{\int_V\romd^3 r \; \rome^{-\beta z_i e \, \psi_\romtot(\VECr)}}.
\end{eqnarray}
Second, the total charge density $e\sum_i z_i n_i(\VECr)$ has to
satisfy Poisson's equation.  This situation arises if the counterions
form a mixture of different valences or if the system contains
additional salt. In this section we would like to make a few remarks
about the latter case.

The amount of salt can be specified by the number of salt molecules
per cell. Since salt molecules are neutral (unlike counterions), their
number is not restricted by the constraint of electroneutrality, and
different cells may contain different numbers of salt molecules. This
variation cannot be taken into account by a description focusing on
just one cell and is thus neglected. One assumes instead that the cell
contains a number of salt molecules equal to the average salt
concentration in the polyelectrolyte solution times the cell
volume. In other words: The division of salt between the cells is
assumed to be perfectly even.

\begin{figure}
  \begin{center}
    \includegraphics[scale=0.6]{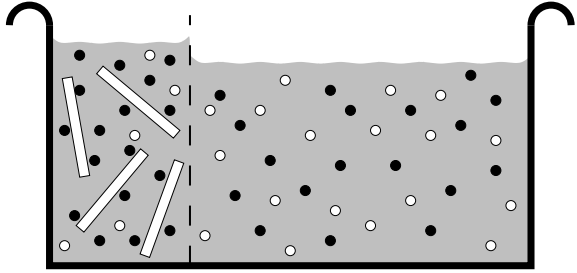}
  \end{center}
  \caption{Solution of charged rod-like polyelectrolytes and counterions
    in equilibrium with a bulk salt reservoir. The membrane is permeable
    for everything but the macroions. Notice that both compartments
    are charge neutral.}
  \label{fig:donnan_cell}
\end{figure}

If the presence of salt is due to the fact that the polyelectrolyte
solution is in contact with a salt reservoir, there is a further
problem to solve: How is the average concentration of salt molecules
in the polyelectrolyte solution related to the concentration of salt
molecules in the salt reservoir? This situation is depicted in
Fig.~\ref{fig:donnan_cell} and is referred to as a ``Donnan
equilibrium'' \cite{Don24,Ove56}. In the following we will address
this question on the level of the cell model and Poisson-Boltzmann
theory. For simplicity we will only treat the case in which all ions
are monovalent.

The compartment containing the macroions will be described by a
cell-model, which apart from the central rod and the counterions will
contain a certain number of salt molecules, yet to be determined.
Since both the counterions and the oppositely charged coions can cross
the membrane, they have to be in electrochemical equilibrium. However,
before we can write down a condition for that, there is an additional
effect that we have to take into account: Since the charged macroions
cannot leave their compartment, their counterions also will have to
remain there for reasons of electroneutrality. Hence, upon addition of
salt there is a tendency for the salt to go into the other
compartment, which is less ``crowded''. However, this implies that in
general there must be a discontinuity in the counter- and coion
density across the membrane. Such a difference can only be sustained
by a corresponding drop in the electrostatic potential across the
membrane separating the two compartments. This potential drop is
referred to as the ``Donnan potential'', $\Phi_\romD$, and must be
taken into account when balancing the electrochemical potentials.

Having said this, we can now proceed to compute electrochemical
potentials on both sides.  On the side containing the macroions we
will compute the chemical potential at the cell boundary, where the
only contribution is the entropy term if we set the potential to zero
there. In the bulk salt reservoir we get an entropy term corresponding
to the bulk salt density, which we call $n_\romb$, and a term
corresponding to the Donnan potential $\Phi_\romD$. With $n_\pm$ being
the cation and anion densities at the cell boundary, we obtain
\begin{eqnarray}
  k_\romB T \, \ln n_\pm
    & = &
  k_\romB T \, \ln n_\romb \; \pm \; e \, \Phi_\romD
  \nonumber \\
  \Rightarrow \qquad n_\pm
    & = &
  n_\romb \, \rome^{\pm \beta e \Phi_\romD}.
  \label{eq:n+-}
\end{eqnarray}
Multiplying cation and anion density gives
\begin{eqnarray}
  n_+ \; n_- \; = \; n_\romb^2.
  \label{eq:n+n-nb2}
\end{eqnarray}
That is, the bulk salt density is the geometric average of the cation
and anion densities at the cell boundary. Dividing the ion densities
in Eqn.~(\ref{eq:n+-}) yields an expression for the Donnan potential:
\begin{eqnarray}
  \beta e \, \Phi_\romD
    \; = \;
  \frac{1}{2}\,\ln\frac{n_+}{n_-}
    \; \stackrel{\text{(\ref{eq:n+n-nb2})}}{=} \;
  \ln\frac{n_+}{n_\romb}.
  \label{eq:phi_Donnan}
\end{eqnarray}
This also shows that the Donnan potential diverges in the zero salt
limit.

For sufficiently dilute solutions the osmotic pressure follows from
the van't Hoff equation $\beta P = \text{[solute]}$. Since the {\em
excess\/} osmotic pressure is given by the difference between the
osmotic pressures at the cell boundary acting from inside and from
outside, we find
\begin{eqnarray}
  \beta P
    \; = \;
  n_+ + n_- - 2 n_\romb
    \; \stackrel{\text{(\ref{eq:n+n-nb2})}}{=} \;
  \big(\sqrt{n_+}-\sqrt{n_-}\big)^2
    \; \ge \;
  0.
  \label{eq:P_Donnan}
\end{eqnarray}

Let $\delta n_+=n_+-n_\romb$ denote the difference between the cation
density at the outer cell boundary and the cation density in the bulk
salt reservoir. Combining Eqns.~(\ref{eq:n+-}), (\ref{eq:phi_Donnan})
and (\ref{eq:P_Donnan}), we can rewrite the pressure as
\begin{eqnarray}
  \frac{\beta P}{n_\romb}
    & = &
  \frac{\big(\sqrt{n_+}-\sqrt{n_-}\big)^2}{n_\romb}
    \;\; = \;\;
  \Big(\rome^{\beta e \Phi_\romD/2} - \rome^{-\beta e \Phi_\romD/2}\Big)^2
    \nonumber \\
    & = &
  4 \, \sinh^2\Big(\frac{1}{2}\ln\frac{n_+}{n_\romb}\Big)
    \;\; = \;\;
  4 \, \sinh^2\Big[\frac{1}{2}\ln\Big(1+\frac{\delta n_+}{n_\romb}\Big)\Big]
    \nonumber \\
    & \stackrel{\delta n_+ \ll n_\romb}{\approx} &
  \Big(\frac{\delta n_+}{n_\romb}\Big)^2.
  \label{eq:P_app_deltan}
\end{eqnarray}

If one wishes to determine the osmotic pressure, one has to solve the
nonlinear Poisson-Boltzmann equation in the presence of salt, subject
to the constraint in Eqn.~(\ref{eq:n+n-nb2}). However, no analytical
solution is known for this case. A numerical solution can be obtained
in the following way: First ``guess'' an initial amount of salt to be
present in the cell, solve the PB equation\footnote{Two simple ways
for achieving this are described in Ref.~\cite{AlCh84}. Although these
authors treat the spherical case, their approach works equally well
for cylindrical symmetry, since the only important point is that the
problem is one-dimensional.}, compute the bulk salt concentration
implied by this amount via Eqn.~(\ref{eq:n+n-nb2}), adjust the salt
content, and iterate until self-consistency is achieved.

An approximate treatment of the problem can be obtained from the
following two assumptions:
\begin{enumerate}
\item{If the amount of salt is small, it may not significantly disturb the
counterion profile from the salt-free case. Hence, one may hope that
the {\em counterion\/} concentration $n_R$ at the cell boundary is
still given by its value from {\em salt-free\/} Poisson-Boltzmann theory.}
\item{At the outer cell boundary the densities of additional
cations and anions due to the salt are equal, say $\Delta n_\roms$.}
\end{enumerate}
Given these two assumptions, the equilibrium condition (\ref{eq:n+n-nb2})
requires
\begin{eqnarray}
  (n_R+\Delta n_\roms)\,\Delta n_\roms \; \approx \; n_\romb^2
  \qquad\Rightarrow\qquad
  2 \Delta n_\roms \; \approx \;
  \sqrt{n_R^2 + (2n_\romb)^2} - n_R.
\end{eqnarray}
Together with Eqn.~(\ref{eq:P_Donnan}) this gives the following
approximate expression for the excess osmotic pressure:
\begin{eqnarray}
  \beta P \; \approx \;
  n_R + 2 \Delta n_\roms -2 n_\romb \; = \;
  \sqrt{n_R^2 + (2n_\romb)^2} - 2n_\romb.
  \label{eq:P_app}
\end{eqnarray}
The advantage of this expression is that it requires only the
knowledge of the counterion density $n_R$ at the outer cell
boundary from salt-free Poisson-Boltzmann theory, which is much
easier to determine than the solution including the salt
explicitly.\footnote{It requires only the numerical solution of
the transcendental equation (\ref{eq:PB_gamma}), not the numerical
solution of a nonlinear differential equation.}

\begin{figure}[t]
  \begin{center}
\begingroup%
  \makeatletter%
  \newcommand{\GNUPLOTspecial}{%
    \@sanitize\catcode`\%=14\relax\special}%
  \setlength{\unitlength}{0.1bp}%
\begin{picture}(3600,2160)(0,0)%
\special{psfile=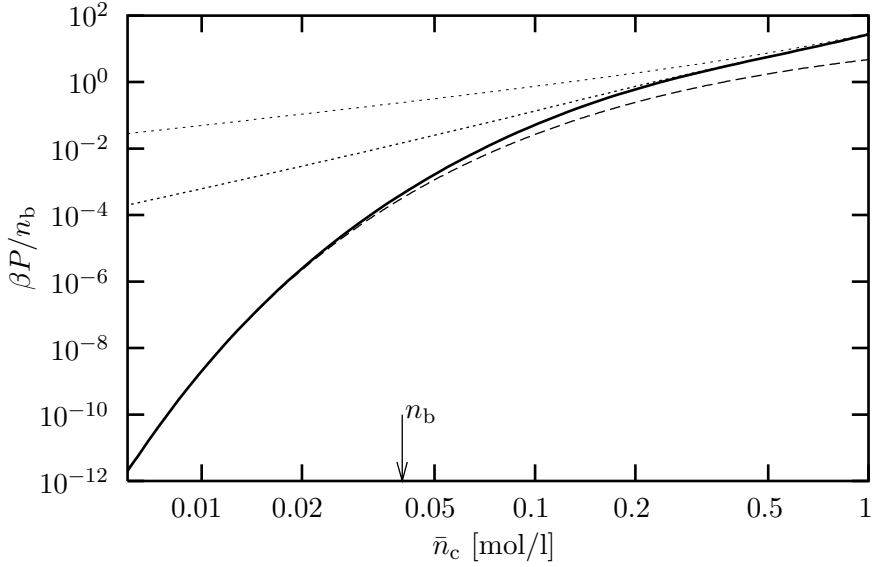 llx=-2 lly=0 urx=720 ury=504 rwi=7200}
\put(1715,551){\makebox(0,0)[l]{$n_\romb$}}%
\put(2050,50){\makebox(0,0){$\bar{n}_\romc$ [mol/l]}}%
\put(300,1180){%
\special{ps: gsave currentpoint currentpoint translate
270 rotate neg exch neg exch translate}%
\makebox(0,0)[b]{\shortstack{$\beta P / n_\romb$}}%
\special{ps: currentpoint grestore moveto}%
}%
\put(3450,200){\makebox(0,0){1}}%
\put(3071,200){\makebox(0,0){0.5}}%
\put(2569,200){\makebox(0,0){0.2}}%
\put(2190,200){\makebox(0,0){0.1}}%
\put(1810,200){\makebox(0,0){0.05}}%
\put(1309,200){\makebox(0,0){0.02}}%
\put(930,200){\makebox(0,0){0.01}}%
\put(600,2060){\makebox(0,0)[r]{$10^{2}$}}%
\put(600,1809){\makebox(0,0)[r]{$10^{0}$}}%
\put(600,1557){\makebox(0,0)[r]{$10^{-2}$}}%
\put(600,1306){\makebox(0,0)[r]{$10^{-4}$}}%
\put(600,1054){\makebox(0,0)[r]{$10^{-6}$}}%
\put(600,803){\makebox(0,0)[r]{$10^{-8}$}}%
\put(600,551){\makebox(0,0)[r]{$10^{-10}$}}%
\put(600,300){\makebox(0,0)[r]{$10^{-12}$}}%
\end{picture}%
\endgroup
 
  \end{center}
  \caption{\small Ionic contribution to the osmotic pressure $\beta P$
  of a DNA solution divided by bulk salt concentration $n_\romb$ as a
  function of average counterion concentration $\bar{n}_\romc$. The
  bulk salt concentration is $n_\romb=40\,\text{mmol/l}$. The solid
  line is the prediction of the Poisson-Boltzmann equation (taking
  into account counterions and salt), the (upper) fine dashed line is
  from Poisson-Boltzmann theory without salt.  The
  short dashed line is the approximation from Eqn.~(\ref{eq:P_app})
  and the long dashed line is a fit to Eqn.~(\ref{eq:Pfit}) within the
  range $\bar{n}_\romc = 6 - 10 \,
  \text{mmol/l}$.}\label{fig:don_P}
\end{figure}

The above approximation is good if the counterion concentration is
large compared to the salt concentration. In the opposite limit of
excess salt concentration Eqn.~(\ref{eq:P_app}) behaves
asymptotically like
\begin{eqnarray}
  \frac{\beta P}{n_\romb}
    \; = \;
  2 \, \bigg[\sqrt{1+\Big(\frac{n_R}%
    {2n_\romb}\Big)^2}-1\bigg]
    \; \approx \;
  2 \, \bigg[1 + \frac{1}{2} \Big(\frac{n_R}%
    {2n_\romb}\Big)^2 - 1\bigg]
    \; = \;
  \Big(\frac{n_R}{2n_\romb}\Big)^2.
  \label{eq:Papproxgoessquare}
\end{eqnarray}
Since $n_R$ is given by $\bar{n}_\romc$ times the osmotic coefficient,
which does not strongly vary with density, this equation implies that
in the salt dominated case the osmotic pressure varies quadratically
with the average counterion concentration. However, this is not born
out by a numerical solution of the Poisson-Boltzmann equation with
added salt, which shows an exponential behavior (see
Fig.~\ref{fig:don_P}). The latter can be understood by the following
simple argument: For large salt content the counterion and coion
density profiles can be expected to merge exponentially with a bulk
Debye-H\"uckel screening constant $\kappa_\romD =\sqrt{8\pi\ell_\romB
n_\romb}$. We may thus assume that
\begin{eqnarray}
  \delta n_+ \; \propto \; n_\romb \, \rome^{-\kappa_\romD R}.
\end{eqnarray}
Since the cell radius is related to the average counterion
concentration via $\bar{n}_\romc = \tilde{\lambda}/\pi R^2$, we can
write together with Eqn.~(\ref{eq:P_app_deltan})
\begin{eqnarray}
  \frac{\beta P}{n_\romb}
    \; \propto \;
  \exp\{-2\kappa_\romD R\}
    \; = \;
  \exp\Big\{-2\sqrt{8\pi\ell_\romB n_\romb}
  \sqrt{\tilde{\lambda}/\pi\bar{n}_\romc}\Big\}.
\end{eqnarray}
Taking the logarithm we finally see
\begin{eqnarray}
  \log\bigg(\frac{\beta P}{n_\romb}\bigg)
    \; = \;
  C_1 - C_2\,\sqrt{\frac{n_\romb}{\bar{n}_\romc}}
  \label{eq:Pfit}
\end{eqnarray}
with some constants $C_1$ and $C_2$.

This functional dependence should hold whenever
$n_\romb/\bar{n}_\romc\gg 1$. However, it also demonstrates that
the range of validity of the cell model reaches its limit for high
salt concentrations. As increasingly more salt is added to the
system the osmotic pressure of the ions vanishes exponentially.
However, this does not imply that the total osmotic pressure of
the polyelectrolyte solution vanishes, since we have neglected the
contribution coming from the macroions \cite{RasCon+00}. Since the
latter generally depends on observables which are largely
irrelevant for the cell model (\eg, the degree of polymerization),
this model must break down here.


\section{Outlook}

We presented in our brief introduction two of the most common
approximations encountered in the theory of charged macromolecules:
The cell-model and Poisson-Boltzmann theory. Where can one go from
here?

One of the key deficiencies of Poisson-Boltzmann theory is the neglect
of interparticle correlations. We have seen how this arose from the
assumption of a product state, which subsequently led to a simple
(local) density functional theory. An important theorem originally due
to Hohenberg and Kohn states that there actually {\em exists\/} a
density functional which gives the correct free energy of the full
system and which differs from the Poisson-Boltzmann functional by an
additional term that takes into account the effects of
correlations.\footnote{A good introduction into this topic can be
found in Ref.~\cite{HaMc86}.} Although the theorem does not specify
what this functional looks like, it shows that attempts that go beyond
Poisson-Boltzmann theory but stay on a density functional level are
not futile. Indeed, various local \cite{DFT:local} and nonlocal
\cite{DFT:nonlocal} corrections to the Poisson-Boltzmann functional
have been suggested in the past.

The Coulomb problem has been treated in a field-theoretic way,
\ie, the classical partition function is transformed via a
Hubbard-Stratonovich transformation into a functional integral
\cite{field}.  Poisson-Boltzmann theory is rediscovered as the
saddle-point of this field theory, and higher order corrections
can in principle be obtained using the large and well-established
toolbox of field-theoretic perturbation theory. There also exists
the possibility to approximate the functional integral in the
limit opposite to Poisson-Boltzmann theory, when correlations
dominate the system \cite{MorNet00}. All this is thoroughly
discussed in the lecture notes by Moreira and Netz
\cite{MoNe:LesHouches}.

A different route to incorporate correlations is offered by
integral equation theories. Their key idea is to first derive
exact equations for various correlation functions and then
introduce some approximate relations between them, based for
instance on perturbation expansions, which lead to integral
equations that implicitly give the desired correlation functions.
This approach and its relation to Poisson-Boltzmann and
Debye-H\"uckel theory is the topic of the lecture of Kjellander
\cite{Kjel:LesHouches}.

A further method for dealing with correlations is to simulate the
systems on a computer and explicitly keep track of all the ions---or
even solvent molecules. This approach has become an increasingly
important tool for both describing real systems as well as testing
approximate theories. More details can be found in the lectures of
J\"onsson and Wennerstr\"om \cite{JoWe:LesHouches} and Holm and Kremer
\cite{Kr:LesHouches}.

Going beyond the cell model and taking the actual shape of
polyelectrolytes into account is in general an extremely difficult
business. However, a remarkable amount of information can be
obtained by using some (or, better yet, a lot of) physical insight
and writing the free energy as a sum of a few terms which account
for the most relevant physical properties of the system (for
instance the chain elasticity, electrostatic self-energy or
hydrophobic interactions) and possibly some variational
parameters. Since in doing so all prefactors are neglected (it
only matters how one observable ``scales'' with another) these
approaches are known as scaling theories. Joanny gives an
introduction and a few famous applications in his lecture
\cite{Joa:LesHouches}.

\vspace*{1em}

All these approaches reach beyond the cell-model and/or the
Poisson-Boltzmann equation. They boldly go where no mean-field theory
has gone before. However, we believe that in order to appreciate their
efforts it is worthwhile to know where they came from.  It was the
intention of this chapter to provide some of that knowledge.


\section*{Acknowledgments}

M.\ D.\ would like to thank P.\ L.\ Hansen and I.\ Borukhov for
stimulating discussions.




\begin{thebibliography}{99}
\bibitem{GB} The formula (\ref{eq:GibbsBogoliubov}) goes under many
  names, for instance ``Gibbs-\-Bogoliubov-\-in\-equality'' \index{Gibbs-Bogoliubov-inequality} or ``Feynman
  variational principle''. Its derivation is remarkably simple:
  For any two arbitrary states $p$ and $p_0$ the Gibbs inequality
  $\romTr\,p_0\ln p \le \romTr\,p_0\ln p_0$ holds, since the
  elementary inequality $\ln x \le x-1$ implies
  \begin{eqnarray}
    \romTr\,p_0\ln p - \romTr\,p_0\ln p_0 =
    \romTr\,p_0\ln \frac{p}{p_0} \le
    \romTr\,p_0\Big(\frac{p}{p_0}-1\Big) =
    \romTr\,p - \romTr\,p_0 = 0.
  \end{eqnarray}
  If one inserts an arbitraty $p_0$ and the canonical $p=\rome^{-\beta H}/\,\romTr\,
  \rome^{-\beta H}$, one obtains
  \begin{eqnarray}
    S_0 & \!\! := \!\! & -k_\romB \; \romTr\,p_0\ln p_0
           \; \le \; -k_\romB \; \romTr\,p_0\ln p
           \; = \; -k_\romB \; \romTr \left\{p_0\;\big(\!-\beta H
             - \ln\romTr\,\rome^{-\beta H}\big)\right\}
           \nonumber\\
   {}   & \!\! =  \!\! & \frac{1}{T} \, \big(\romTr\,p_0 H
                         + k_\romB T \ln \romTr\,\rome^{-\beta H}\big)
          \; =: \; \frac{1}{T}\,\big(\langle H \rangle_0 - F\big),
  \end{eqnarray}
  which is formula (\ref{eq:GibbsBogoliubov}). Incidentally, the Gibbs inequality
  remains valid if $p$ and $p_0$ are quantum states --- and so does the
  Gibbs-Bogoliubov-inequality.
\bibitem{RolMan} This has been discussec by R.\ R.\ Netz in his Les Houches
  lecture, but is not included in his contribution in
  Ref.~1.
\end{thebibliography}
\end{document}